\documentclass[sigconf]{acmart-me}
\usepackage{booktabs}
\usepackage{url}
\usepackage{color}
\usepackage{enumitem}
\setcopyright{rightsretained}

\acmDOI{}

\acmISBN{}

\acmConference[MediaEval'20]{Multimedia Evaluation Workshop}{14-15 December 2020}{Online}
\acmYear{2020}
\copyrightyear{2020}
\acmPrice{}

\begin{document}
\title{Leveraging Audio Gestalt to Predict Media Memorability}
\author{Lorin Sweeney, Graham Healy, Alan F. Smeaton}
\affiliation{Insight Centre for Data Analytics, Dublin City University, Glasnevin, Dublin 9, Ireland\\}
\email{lorin.sweeney8@mail.dcu.ie}

\renewcommand{\shortauthors}{L. Sweeney, G. Healy, A. Smeaton}
\renewcommand{\shorttitle}{Predicting Media Memorability}

\begin{abstract}
Memorability determines what evanesces into emptiness, and what worms its way into the deepest furrows of our minds. It is the key to curating more meaningful media content as we wade through daily digital torrents. The Predicting Media Memorability task in MediaEval 2020 aims to address the question of media memorability by setting the task of automatically predicting video memorability. Our approach is a multimodal deep learning-based late fusion that combines visual, semantic, and auditory features. We used audio gestalt to estimate the influence of the audio modality on overall video memorability, and accordingly inform which combination of features would best predict a given video's memorability scores.
\end{abstract}
\maketitle

\section{Introduction and Related Work}
\label{sec:intro}
Our memories make us who we are---holding together the very fabric of our being. The vast majority of the population predominantly rely on visual information to remember and identify people, places, and things \cite{kirkpatrick1894}. It is known that people generally tend to remember and forget the same images \cite{1,2}, which implies that there are intrinsic qualities or characteristics that make visual content more or less memorable. While there is some evidence to suggest that sounds similarly have such intrinsic properties \cite{soundMem}, it is generally accepted that auditory memory is inferior to visual memory, and decays more quickly \cite{bigelow2014,cohen2009,backman1998}. However, it is important not to resultantly resort to dismissing the role of the audio modality in memorability---as multisensory experiences exhibit increased recall accuracy compared to unisensory ones \cite{thelen2015,thelen2013}---but to avail of the potential contextual priming information sounds provide \cite{schirmer2011}. In our work, we seek to explore the influence of the audio modality on overall video memorability, and employ \cite{soundMem}'s concept of audio gestalt in order to do so. 

Image memorability is commonly defined as the probability of an observer detecting a repeated image in a stream of images a few minutes after exposition \cite{1,2,3,4}. This paper outlines our participation in the 2020 MediaEval Predicting Media Memorability Task \cite{5} where memorability is sub-categorised into short-term (after minutes) and long-term (after 24-72 hours) memorability. The task requires participants to create systems that can predict the short-term and long-term memorability of a set of viral videos. The dataset, annotation protocol, pre-computed features, and ground-truth data are described in the overview paper \cite{5}.
\begin{figure}
\includegraphics[width=0.8\columnwidth]{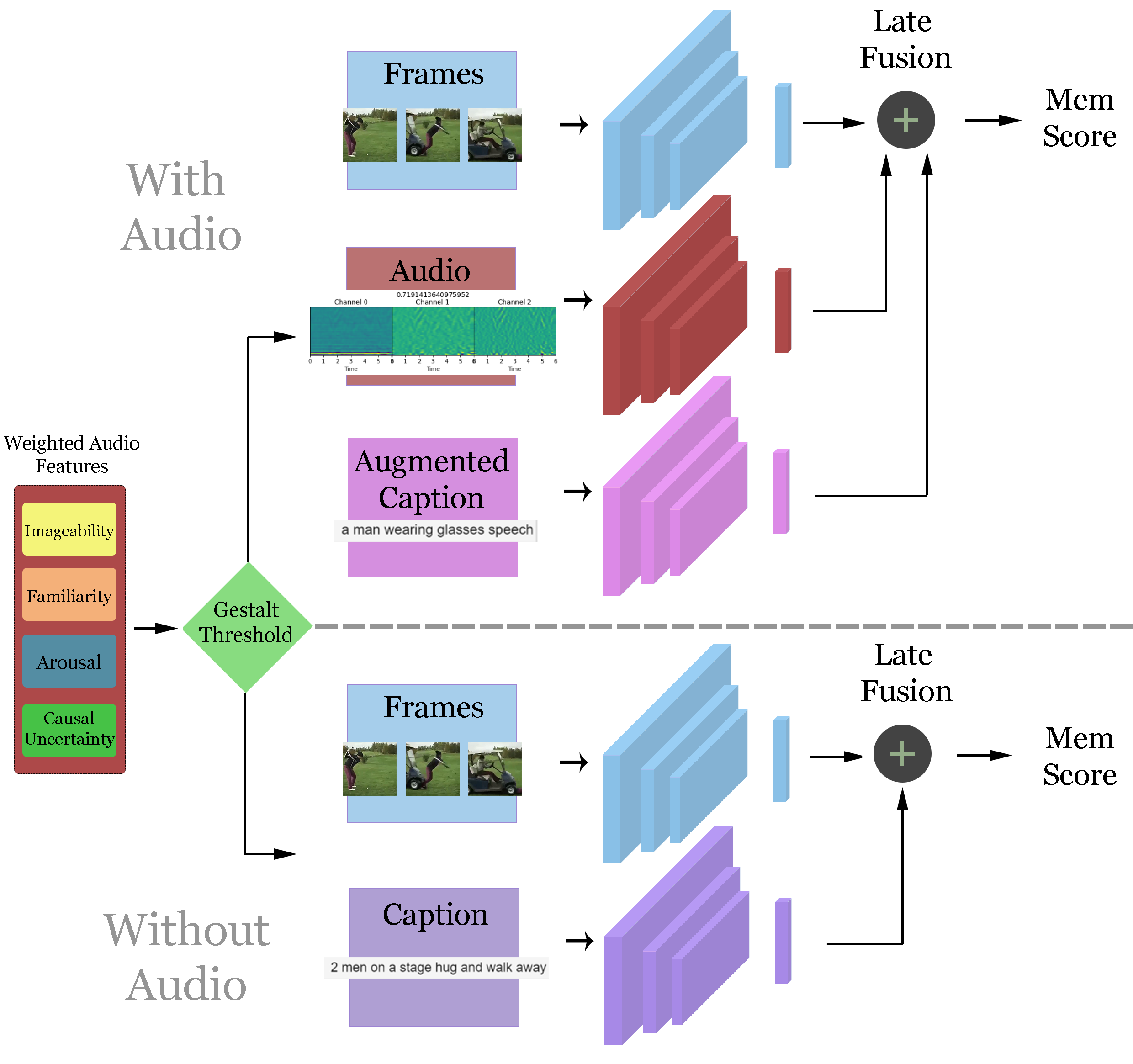}
\caption{Our multimodal deep-learning based late fusion framework,  using conditional audio gestalt based threshold}
\end{figure}
Previous attempts at computing memorability have consistently demonstrated that off-the-shelf pre-computed features, such as C3D; HMP; LBP; etc., have mostly been unfruitful \cite{6,7,8,9}, with captions being the exception \cite{10,11}. The high usefulness of captions is likely due to the semantically rich nature of text---the medium with the highest cued recall and free recall for narrative \cite{furnham1990}---capturing the most information about the elements in the videos with the smallest semantic gap. Recent attempts highlighted the effectiveness of deep features in conjunction with other semantically rich features, such as emotions or actions \cite{12,13}. To our knowledge, the influence of audio on overall video memorability has yet to be explored.

\section{Our Approach}
\label{sec:approach}
The dataset is comprised of three sets---an initial development set, a supplemental development set, and a test set. Both development sets contain ground-truth (GT) scores and anonymised user annotation data for 590 and 410 videos respectively, while the test set contains 500 videos without GT scores. Upon inspection of the supplemental development set, we noticed that the short-term GT scores were true-negative rates (probabilities that unseen videos would not be falsely remembered), rather than true-positive rates, like the rest of the scores. We accordingly decided to generate our own short-term GT scores by employing collaborative filtering with the provided reaction time annotations. The resulting matrix of predicted reaction times, for each user-video combination, was used to calculate a short-term memorability score for each video. Predicted reaction times that were more than two standard deviations from the mean reaction time were counted as misses. We then divided this new development set into a training (800 videos) and testing set (200 videos). Our approach is predicated on the conditional exclusion/inclusion of audio related features depending on audio gestalt levels.

\textbf{Audio Gestalt:}
Defined in \cite{soundMem} as high-level conceptual audio features, imageability; human causal uncertainty (Hcu); arousal; and familiarity, were found to be strongly correlated with audio memorability. We predict audio gestalt by doing a weighted sum of these four features. Imageability is based on whether the audio is music or not using the PANNs \cite{panns} network. Hcu and arousal scores are predicted with an xResNet34 pre-trained on ImageNet \cite{imageNet}, and fine-tuned on HCU400 \cite{HCU400}. For familiarity we chose to use the top audio-tag confidence score of the PANNs \cite{panns} network, as we had observed a correlation between the two scores. These scores were then weighted and summed to produce an audio gestalt score for a video. Depending on a threshold, one of two pathways---without audio, using captions and frames, and with audio, using audio augmented captions, frames, and audio spectrograms---was used to predict memorability scores.

\textbf{Without Audio:}
Deep Neural Network (DNN) frame-based and caption-based models were chosen, as they had proven to be quite effective in previous video memorability prediction attempts \cite{10,11}. For our caption model, given that overfitting was a primary concern, we decided to use the AWD-LSTM (ASGD Weight-Dropped LSTM) architecture \cite{14}, as it is highly regularised, and is used in state-of-the-art language modelling. In order to fully take advantage of the high level representations that a language model offers, we used transfer learning. The specific transfer learning method employed was UMLFiT \cite{15}, a method that uses discriminative fine-tuning, slanted triangular learning rates, and gradual unfreezing to avoid catastrophic forgetting. A language model was pre-trained on the Wiki-103 dataset, and fine-tuned on the first 300,000 captions from Google's Conceptual Captions dataset \cite{GCC}. The encoder from that fine-tuned language model was then re-used in another model of the same architecture, but trained on 800 of the development set captions to predict short-term and long-term memorability scores rather than the next word in a sentence. For our frame based model, we transfer trained an xResNet50 model pre-trained on ImageNet \cite{imageNet}. It was first fine-tuned on Memento10k \cite{mem10k}, then fine-tuned on 800 of the development set videos.

\textbf{With Audio:}
We incorporated audio features by augmenting captions with audio tags, and training a CNN on audio spectrograms. We kept the same frame based model as a control. We opted to augment captions with audio tags, rather than relying on audio tags alone, so that the context provided by captions was not lost. We used the PANNs model \cite{panns} to extract audio tags, append them to their corresponding development set captions, and re-trained the aforementioned caption model. For the audio spectrogram model, we extracted Mel-frequency cepstral coefficients from the audio, and stacked them with their delta coefficients in order to create three channel spectrogram images. We then used the same transfer training procedure as our frame based model.

For each stream, final predictions were the result of a weighted sum of their constituent model predictions.

\section{Discussion and Outlook}
Table 1 shows the scores for our runs when tested on 200 dev set videos kept for validation. Runs 1-3 were intended as controls for our 4\textsuperscript{th} run in which we tested our proposed framework. Run 1 was the ``audio only'' control, using augmented captions and audio spectrograms; run 2 was the ``no audio'' control, using captions and video frames; and run 3 was the ``everything'' control, using all features irrespective of audio gestalt scores. These preliminary result suggest that the audio gestalt-based conditional inclusion of audio features does indeed improve memorability prediction (run 3 vs run 4).

Table 2 shows the scores achieved by each of our submitted runs. Unfortunately, due to a mistake in the submission process, we do not have official short-term results for our 3\textsuperscript{rd} run, and are therefore unable to properly compare it with run 4 and validate our preliminary results. The stark difference between the official results and our preliminary results highlight the inherent difficulty of achieving generalisability when learning from limited data. Unfortunately, there is no way yet of knowing if these differences are simply due the lack of distributional overlap between training and testing videos, or if it is a case of overfitting.
Surprisingly, our highest short-term score was achieved by a model not trained on any data from the task---an xResNet50 model pre-trained on ImageNet \cite{imageNet}, and fine-tuned on Memento10k \cite{mem10k}. We believe that this result is most likely due to the better generalisability of models trained on larger and more diverse datasets. Our highest long-term score was achieved by our ``audio only'' run, which indicates that the audio modality does indeed provide some contextual information during video memorability recognition tasks.

Further investigation into the influence of the audio modality on overall video memorability is clearly required. Testing our proposed framework on a much larger video memorability dataset would be an interesting next step towards that goal.

\begin{table}[t]
\caption{Results on 200 dev-set videos kept for validation for each of our runs.}
\label{tab:results2}
\begin{tabular}{cccc}
    \toprule
     &\multicolumn{1}{c}{Short-term} && \multicolumn{1}{c}{Long-term} \\\cline{2-4}
    \textbf{Run} & \textbf{Spearman}     && \textbf{Spearman}  \\
    run1-required & 0.345 && 0.365  \\
    run2-required & 0.338  && 0.437  \\
    run3-required & 0.319 && 0.425  \\
    run4-required & \textbf{0.364}   && \textbf{0.470}  \\
    memento10k & 0.314  && -  \\
    \bottomrule
\end{tabular}
\end{table}
\begin{table}[t]
\caption{Official results on test-set for each of our submitted runs.}
\label{tab:results1}
\hspace*{-0.2cm}
\begin{tabular}{cccccccc}
    \toprule
     &\multicolumn{2}{c}{Short-term} && \multicolumn{2}{c}{Long-term} \\\cline{2-3}\cline{5-6}
    \textbf{Run} & \textbf{Spearman } & Pearson     && \textbf{Spearman } & Pearson \\
    run1-required & 0.054 & 0.044 && \textbf{0.113} & 0.121 \\
    run2-required & 0.05 & 0.072  && 0.059 & 0.071 \\
    run3-required & - & - && 0.109 & 0.119  \\
    run4-required & 0.076 & 0.092  && 0.041 & 0.058  \\
    memento10k & \textbf{0.137} & 0.13  && - & - \\
    \bottomrule
\end{tabular}
\end{table}

\subsection*{Acknowledgements}
This publication has emanated from research supported by Science Foundation Ireland (SFI) under Grant Number SFI/12/RC/2289\_P2, co-funded by the European Regional Development Fund.

\bibliographystyle{ACM-Reference-Format}
\def\bibfont{\small} 
\bibliography{sigproc} 

\end{document}